\documentstyle[12pt]{article}

\begin{document}

\begin{titlepage}

\begin{flushright}
IJS-TP-96/27\\
DTP/96/104\\
NUHEP-TH-96-8\\
December 1996\\
\end{flushright}

\vspace{.5cm}

\begin{center}
{\Large \bf $SU(3)$ flavor symmetry breaking and semileptonic decays 
of heavy mesons}

\vspace{1.5cm}

{\large \bf B. Bajc $^{a,b}$
\footnote{British Royal Society 
Postdoctoral Fellow}, S. Fajfer $^{a,c}$ and
Robert J. Oakes $^{c}$}

\vspace{.5cm}

{\it a) J. Stefan Institute, Jamova 39, P.O. Box 3000, 
1001 Ljubljana, Slovenia}

\vspace{.5cm}

{\it b) Department of Physics, University of Durham, 
Durham DH1 3LE, Great Britain}

\vspace{.5cm}

{\it c) Department of Physics and Astronomy, 
Northwestern University, Evanston, Il 60208
U.S.A.}

\vspace{1cm}

\end{center}

\centerline{\large \bf ABSTRACT}

\vspace{0.5cm}

Assuming the ${\bar D}^0$, $D^-$, $D^-_s$ and 
$B^+$, $B^0$, $B_s^0$ mesons belong to 
triplets of $SU(3)$ flavor symmetry, we analyse the form factors in the 
semileptonic decays of these mesons.
Both quark and meson mass differences are taken into account. 
We find a number of relations, in agreement with the present data 
as well as with previous analyses, and predict certain ratios of 
form factors, not yet measured, most notably the D meson decay  
constant $f_D = 209 \pm 39$ MeV.

\noindent
PACS numbers(s): 13.20.-v, 13.20.Fc, 13.20.He \\

\end{titlepage}

\setlength {\baselineskip}{0.75truecm}

The B meson semileptonic decays are of special importance in our 
understanding of the Standard Model (see e.g. \cite{GM,AA}). 
Many theoretical approaches have been developed to describe 
these decays: heavy quark effective  theory (HQET) 
\cite{BG}-\cite{G2}, lattice calculations;   
%(need ref.s)
e.g., \cite{GM,LUB,DAMIR}, QCD sum rules; e.g., \cite{PB}-\cite{KR}, 
the quark-hadron duality model \cite{ACCM}, 
and other quark models \cite{ISGW}-\cite{KS}.

The heavy quark symmetry \cite{IW}, \cite{HG}
relates the dynamical variables of  
D meson decays with the decay variables of B mesons. 
Using the heavy quark 
symmetry for the  $c$ and $b$ quarks, the 
ratio of decay constants $(f_{B_s}/f_{B_d})/(f_{D_s}/f_{D})$ 
and the ratio of  the semileptonic form factors 
$(f_+^{B\pi}/f_+^{BK})/(f_+^{D\pi}/f_+^{DK})$
in B and D meson decays has been  estimated \cite{G}. 
This approach is limited by the fact that charm and bottom quark   
masses, $m_c$ and $m_b$,
are, indeed, finite and there are  
corrections to the results of HQET 
that  are of the order $m_s/m_c$ \cite{G,RJO}.
Moreover, 
the predictions of heavy quark effective theory are 
expected to be reliable only
in the region of small recoil momentum ($q^2\approx q^2_{max}$).

On the experimental side, there are many new 
results in D leptonic and semileptonic decays \cite{RPD}.
The CLEO experiment \cite{CLEO1} has  
measured the ratio $|f_+^{D\pi} (0)/f_+^{DK}(0)|= 
1.29\pm0.21\pm 0.11$, assuming  
pole-dominance behavior of the form factors. 
The decay width for $D_s \to \mu \nu_{\mu}$ 
has now also been measured \cite{RPD}. 
The accumulating experimental data \cite{RPD} 
indicate that  there is a 
timely need to better understand the 
behavior of semileptonic decay form factors at  $q^2 \simeq 0$.

Recently \cite{RJO}, the ratios $f_{B_s}/f_{B}$ and $f_{D_s}/f_{D}$, 
as well as the double ratio $(f_{B_s}/f_{B})$$/(f_{D_s}/f_{D})$ 
have been analysed, following the 
usual assumption that chiral symmetry is broken by quark mass 
terms in the energy density, and assuming the 
heavy mesons ${\bar D}^0, D^-, D_s^-$ and  $B^+, B^0, B_s^0$ 
belong to triplets of the $SU(3)$ flavor group; i.e., assuming 
the {\it state vectors} of these mesons in their 
rest frame are not appreciable mixed with other representations 
of $SU(3)$.
Following the analysis of ref. \cite{RJO}, 
and relying on the current experimental data,  
we analyse the effects  
of the $SU(3)$ symmetry breaking, due to the substantial quark 
and meson mass splittings, 
in semileptonic decays of heavy pseudoscalar 
mesons into light pseudoscalar mesons.

We use the usual definitions

\begin{equation}
\label{e1}
<0 | A_{\mu} (x) | H(p) > = i f_H p_{\mu} e^{- i p \cdot x}\;,
\end{equation}

\noindent
and 

\begin{equation}
\label{e2}
<P(p^{\prime}) | V_{\mu} (x) | H(p) > = 
[f_+(q^2) (p_{\mu} + p^{\prime}_{\mu}) +
f_-(q^2) (p_{\mu} - p^{\prime}_{\mu}) ]
e^{- i (p - p^{\prime})\cdot x}\;,
\end{equation}

\noindent
where $H(p)$ is generic for one of the heavy mesons 
${\bar D}^0, D^-, D_s^-$ or  $B^+, B^0, B_s^0$ and 
$P(p^{\prime})$ denotes one of  the light 
pseudoscalar states $\pi, K, \eta$ or $\eta'$, and   
$q = p - p^{\prime}$.
The vector and axial-vector current operators are 
$V_{\mu} = {\bar Q} (x) \gamma_{\mu} q(x)$
and 
$A_{\mu} = {\bar Q} (x) \gamma_{\mu} \gamma_5 q(x)$ with $q(x)$ and 
$Q(x)$ 
being the light and heavy quark field operators, respectively. 
We assume, as usual, that chiral symmetry is broken only by 
quark mass terms. The energy density is then of the form $H_0(x)+ H_1(x)$,
where $H_0$ is chiral symmetric and $H_1$ breaks this symmetry. 
In terms of (current) quark masses and fields $H_1$ is given 
by 

\begin{equation} 
\label{e3}
H_1(x) = \sum_i m_i {\bar q}_i (x) q_i(x)\;, 
\end{equation}

\noindent
where the sum extends over all quarks, both light and heavy. 
The divergences of the axial vector and vector currents 
can readily be calculated from (\ref{e3})
using the local relations

\begin{equation} 
\label{e4}
\partial^{\mu} A_{\mu}(x) = -i [Q_5, H_1(x)]\;,
\end{equation}
and 
\begin{equation} 
\label{e5}
\partial^{\mu} V_{\mu}(x) = -i [Q, H_1(x)]\;,
\end{equation}

\noindent
where $Q_5$ and $Q$ are the axial and the vector charges.
One easily finds the matrix elements 

\begin{equation} 
\label{e6}
< 0| \partial^{\mu} A_{\mu}(0)| H(p)> = 
i (m_Q + m_q) <0|{\bar Q} (0) \gamma_5 q(0)| H(p)>\;,
\end{equation}

\noindent
and 

\begin{equation} 
\label{e7}
< P(p^{\prime})| \partial^{\mu} V_{\mu}(0)| H(p)> = 
i (m_Q - m_q) < P(p^{\prime})|{\bar Q} (0)q(0)| H(p)>\;,
\end{equation}

\noindent
which together with (\ref{e1}) and (\ref{e2}) give the exact 
relations 

\begin{equation} 
\label{e9a}
f_H m_H^2 = 
i (m_Q + m_q) < 0|{\bar Q} \gamma_5 q| H(p)>\;,
\end{equation}

\noindent
and 

\begin{equation} 
\label{e9v}
f_+(q^2) (m_H^2 - m_P^2) + f_-(q^2) q^2 = 
- (m_Q - m_q) < P(p^{\prime})|{\bar Q} q| H(p)>\;.
\end{equation}

Proceeding to analyse the constraints 
on the form factors 
coming from (\ref{e9a}) and 
(\ref{e9v}), we first notice that 
the density operators in the matrix elements,  
{\it by definition}, belong to the same 
$SU(3)$ triplet representation. 
Next, we will assume the vacuum state is a singlet under $SU(3)$, while 
the light pseudoscalar mesons belong to the octet representation. 
While neither of these assumptions can be exact, since $SU(3)$ is broken, 
neglecting the mixing of other representations in these {\it state vectors} 
has proven to be an excellent approximation in numerous applications over 
many years \cite{GY}. It has been remarkably well established, 
empirically, that the breaking of light quark $SU(3)$ symmetry 
by only the current quark mass terms in $H_1$, while  producing 
quite  large mass splittings, 
does not appreciably mix the state vectors. 
We will further assume that 
the heavy meson state vectors transform 
as triplets under $SU(3)$. (Actually, it is only the short distance 
parts of the state vectors that need belong to pure $SU(3)$ 
representations, since we will be computing the matrix elements only of
local operators.)  
The Wigner - Eckart theorem can then be applied 
to estimate the ratio of the matrix elements
using the  Clebsch-Gordan coefficients for the 
$3 \times {\bar 3} = 8 + 1$ 
represntations of the $SU(3)$ flavor group. 
Moreover, one can 
expect that this application of the Wigner - Eckart theorem 
to be even better for the heavy mesons since they are 
more compact systems than  the light 
quark systems and $SU(3)$ flavor symmetry becomes exact in the short 
distance limit, where 
the light quark masses can be ignored altogether.

Using the defintion of the scalar form factor $f_0(q^2)$ 
\begin{equation}
\label{e16}
f_- (q^2) = [ - f_+(q^2) + f_0(q^2)] {(m_H^2 -m_P^2)}/{q^2}\;,
\end{equation}

\noindent
one also finds 

\begin{equation}
\label{e17}
(m_H^2-m_P^2) f_0(q^2) = 
- (m_Q - m_q) < P(p^{\prime})| {\bar Q} q| H(p)>\;.
\end{equation}

\noindent
Recall that in the heavy quark limit 
$f_+ = -f_-$ and $f_0 =0$ \cite{BG,G,MN}. 
The effects of $SU(3)$ flavor symmetry breaking  
on semileptonic decays in HQET were discussed in refs. \cite{BG,MN,BD}. 
However, these approaches are expected to be valid only in the
region of small recoil momentum; that is, at $q^2_{max}$, which is not a 
limitation of the present approach. 
We emphasize that (\ref{e9a}) - (\ref{e17}) are exact relations for 
all $q^2$. Consequently, we can analyse the 
form factors in the region $q^2 \simeq 0$, which is more accessible, 
experimentally.

In the following we concentrate on the 
semileptonic decays of $D^0, D^+, D_s^+$ mesons.
Putting $q^2= 0$ in (\ref{e9v}) and taking the ratio of the 
form factors for 
$D \to K l\nu_l$ and $D \to \pi l\nu_l$ one 
obtains 

\begin{equation} 
\label{e10}
\frac{f_+^{DK}(0) (m_D^2 - m_K^2)}{f_+^{D \pi}(0) (m_D^2 - m_{\pi}^2)}= 
 \frac{ m_c - m_s}{m_c - m_u} 
\frac{ < K(p^{\prime})|{\bar s} c| D(p)>}
{< \pi(p^{\prime})|{\bar u} c| D(p)>}|_{q^2=0}\;.
\end{equation}

\noindent
Applying the Wigner-Eckart theorem, as discussed above,  
one  expects that 

\begin{equation}
\label{e11}
\frac{ < K(p^{\prime})|{\bar s}\Gamma c| D(p)>}
{< \pi(p^{\prime})|{\bar u}\Gamma c| D(p)>}|_{q^2=0} \simeq 1
\end{equation} 

\noindent
($\Gamma=1,\gamma_5$) is a very good approximation. 
>From (\ref{e9a}), as in ref. \cite{RJO}, we also have 

\begin{equation}
\label{e12}
\frac{f_{D_s}}{f_D} \frac{m_{D_s}^2}{m_D^2} = 
1 + \frac{m_s}{m_c}\;.
\end{equation}

\noindent
where $m_u/m_c$ has been neglected. From (\ref{e10}) - (\ref{e12}) 
the dependence on quark masses can be eliminated, yeilding the following 
relation among ratios of D meson decay form factors and meson masses:

\begin{equation} 
\label{e13}
\frac{f_+^{DK}(0)}{f_+^{D \pi}(0) }= 
[ 2 - \frac{f_{D_s}}{f_{D}} \frac{m_{D_s}^2}{m_{D}^2}] 
\frac{m_D^2 - m_{\pi}^2}{m_D^2 - m_{K}^2}\;.
\end{equation}

\noindent
In (\ref{e13}) only $f_D$ is not yet experimentally measured and 
therefore it can be used to calculate $f_D$:

\begin{equation} 
\label{e14}
f_D = f_{D_s} \frac{m_{D_s}^2}{m_{D}^2}
[2 - \frac{f_+^{DK}(0)}{f_+^{D \pi}(0) } 
\frac{m_D^2 - m_{K}^2}{m_D^2 - m_{\pi}^2}]^{-1}\;.
\end{equation}

\noindent
The $D_s \to \mu \nu$ decay constant has been measured  to be 
$f_{D_s} = 241 \pm 21\pm30$ MeV \cite{R}.
>From (\ref{e14}) we then predict 

\begin{equation}
\label{e15}
f_D = 209\pm 39 \;\;\mbox{MeV}, 
\end{equation}

\noindent
where we have also used the experimentally measured value 
$|f_+^{D\pi}(0)/f_+^{D K}(0)| = 1.29\pm 0.21 \pm 0.11$ \cite{CLEO1}, 
assuming the positive sign for this ratio.  
This prediction for $f_D$, (\ref{e15}), is in a remarkable 
agreement with the lattice calculations 
\cite{GM}, lending support for our approach. 
(For $f_+^{D\pi}(0)/f_+^{D K}(0) = -1.29$ we obtain the 
rather unlikly value $f_D = 98$ MeV.) 
The error in (\ref{e15}) reflects only  the 
experimental errors on $f_+^{D\pi}(0)/f_+^{D K}(0)$ 
and $f_{D_s}$. 
One can also determine the ratio of the current 
quark masses, $m_s/m_c = 0.27 \pm 0.13$, 
from (\ref{e12}) and (\ref{e13}), 
which also agrees with the data  \cite{RPD}, although the uncertainty is 
large.

We can empirically check the validity of 
our approach using independent data as follows: 
The decay rates for 
$D_s \to \eta (\eta^{\prime}) \mu \nu_{\mu}$ have been measured \cite{RPD}.
If we assume that  the $\eta$ and $\eta^{\prime}$ states mix as usual 
( $\eta = \eta_8 cos \theta - \eta_0 sin \theta$ and
$\eta^{\prime} = \eta_8 sin \theta  + \eta_0 cos\theta$), 
and use (\ref{e9v}), we have

\begin{equation}
\label{e18}
f_+^{D_s \eta}(0) = - {\sqrt \frac{2}{3}} 
(cos \theta  + \frac{1}{{\sqrt 2}} sin\theta) f_+^{DK}(0)\;,
\end{equation}

\noindent
and 

\begin{equation}
\label{e19}
f_+^{D_s {\eta}^{\prime}}(0) = - {\sqrt \frac{2}{3}} 
(sin \theta - \frac{1}{{\sqrt 2}} cos\theta) f_+^{DK}(0)\;.
\end{equation}

\noindent
>From the $D \to K l \nu_l$ decay data 
the form factor $f_+^{DK}(0) = 0.74 \pm 0.03$ \cite{RPD}.
This value was extrapolated from the branching ratio assuming 
pole-dominance. Assuming 
that the form factors $f_+^{D_s \eta 
(\eta^{\prime})}$ are pole-dominated, too, (see, e.g. \cite{WSB1})  
we then also have 

\begin{equation}
\label{e20}
f_+^{D_s \eta (\eta^{\prime)}}(q^2) 
= {f_+^{D_s \eta (\eta^{\prime})}(0)\over
1 - q^2/m_{D^*_s}^2}\;.
\end{equation}

\noindent
Calculating the decay rates for $D_s \to \eta (\eta^{\prime}) \mu \nu_{\mu}$
as in ref. \cite{BFO} one finds 

\begin{equation}
\label{e21}
\Gamma (H \to P l \nu_l) = \frac{G_F^2 m_H^2}{24 \pi^3} 
\int_0^{y_m} dy \frac{|V_{qq^{\prime}} f_+ (0)|^2}{(1 - 
\frac{m_H^2}{m_H^{*2}} y)^2} [ \frac{(m_H^2 (1 - y) + m_P^2)^2}{4 m_H^2} 
-m_P^2]^{3/2}\;,
\end{equation}

\noindent
where $V_{qq^{\prime}}$ denotes the CKM matrix element, and $m_H^*$ is the 
mass of the heavy vector meson causing the 
pole in the form factors.
In Table I we present the branching ratios for the 
$D_s \to \eta (\eta^{\prime}) \mu \nu_{\mu}$ decays, assuming 
the typical values of the mixing angle 
$\theta = -10^0$,$ -20^0$, and $ -23^0$, 
as given in \cite{RPD}. Table I 
indicates 
that $\theta = -10^0$ is 
prefered, although the experimental data are not very precise, which is 
consistent with the 
Gell-Mann-Okubo mass relation \cite{GO}, which lends 
further phenomenological 
support for our approach. 

Finally, we note that 
from (\ref{e16}) it follows that $f_0(0)=f_+(0)$, 
and therefore  (\ref{e13}), (\ref{e18}) and (\ref{e19}) are 
then also valid for the scalar form factor $f_0(0)$.

The semileptonic decays of B mesons can be divided into 
two categories. First are the usual weak decays $B \to \pi l \nu_l$,  
where knowledge of the form factor $f_+$ 
enables one to estimate the CKM matrix element $|V_{ub}|$ 
\cite{MA}. 
The second type is the flavor-changing neutral   
decays $B \to K \mu^+ \mu^-$. The calculation of the rate for 
these decays not only 
requires knowing the form 
factors $f_+^{BK}$ but also the additional 
form factor $h^{BK}$, defined  by 
the matrix element 
$<K(p^{\prime})| {\bar b} \sigma_{\mu \nu} s|B(p)> = -2 ih^{BK}
(p^{\prime}_{\mu} p_{\nu} - p^{\prime}_{\nu} p_{\mu})$. 
These flavor changing neutral decays are   
beyond the scope of the present analysis, but are discussed in 
\cite{BG,G,MN} using HQET. 

To analyse the charged semileptonic decays, we make 
the replacement $c \to b$ in (\ref{e12}), finding  

\begin{equation}
\label{e22}
\frac{f_{B_s}}{f_B} \frac{m_{B_s}^2}{m_B^2} = 1 + \frac{m_s}{m_b}\;.
\end{equation}

\noindent 
Because there are fewer experimental data on B meson decays, 
and since $m_s$ is poorly known, 
we will use the estimate $m_s/m_c = 0.27$, derived from
(\ref{e12}) and (\ref{e13}) above, together with 
the better  established values of the current 
quark masses $m_c = 1.5$ GeV and $m_b = 5$ GeV  
\cite{RPD}. Then,  with $m_s/m_b = 0.081$, we obtain 

\begin{equation}
\label{e23}
\frac{f_{B_s}}{f_B} = 1.04\;,
\end{equation}

\noindent
which is 
very close to the values obtained  previously \cite{GM,G,RJO}. 
Using  (\ref{e13}) with the replacement $c \to b$, 
we also find 

\begin{equation} 
\label{e24}
\frac{f_+^{B K}(0)}{f_+^{B \pi}(0) }= 
[ 2 - \frac{f_{B_s}}{f_{B}} \frac{m_{B_s}^2}{m_{B}^2}] 
\frac{m_B^2 - m_{\pi}^2}{m_B^2 - m_{K}^2}\;,
\end{equation}

\noindent
from which one obtains 

\begin{equation}
\label{e25}
\frac{f_+^{B \pi}(0)}{f_+^{B K}(0)} = 1.08 \;.
\end{equation}

\noindent
Combining this result with the experimental value of 
$|{f_+^{D \pi}(0)}/{f_+^{D K}(0)}|$, the resulting double ratio is 

\begin{equation}
\label{e26}
|{\frac{f_+^{B K}(0)}{f_+^{B \pi}(0)}}/ 
{\frac{f_+^{D K}(0)}{f_+^{D \pi}(0)}}| = 1.19 .
\end{equation}
\noindent
We note that the heavy quark effective theory prediction for this ratio is 
$1$ in the regime of $q^2_{max}$  \cite{G}.

While the majority of the present experimental data 
comes from the decay 
${\bar B}^0 \to \pi^+ l \nu_l$, 
within the antitriplet $B^-$, ${\bar B^0}$, 
and ${\bar B^0_s}$ there are several  
other semileptonic decays to light mesons which are also of interest. 
These other decays also 
offer a possible means to determine the CKM 
matrix elemen $|V_{ub}|$.  
Following our analysis above, with the replacement $c \to b$, 
we have

\begin{equation}
\label{e27}
\frac{f_+^{B^+ \pi^0}(0)}{f_+^{B^0 \pi^-}(0)} = \frac{1}{{\sqrt 2}}\;,
\end{equation}

\begin{equation}
\label{e27a}
\frac{f_+^{B^+ \eta}(0)}{f_+^{B^0 \pi^-}(0)} = \frac{1}{{\sqrt 6}}
(cos \theta - {\sqrt 2} sin \theta)\;,
\end{equation}

\begin{equation}
\label{e27b}
\frac{f_+^{B^+ \eta^{\prime}}(0)}{f_+^{B^0 \pi^-}(0)} = \frac{1}{{\sqrt 6}}
(sin \theta + {\sqrt 2} cos \theta)\;,
\end{equation}

\noindent
and 

\begin{equation}
\label{e27c}
\frac{f_+^{B^0_s K^-}(0)}{f_+^{B^0 \pi^-}(0)} = 
\frac{m_B^2 - m_{\pi}^2}{m_{B_s}^2 - m_K^2}\;.
\end{equation}

\noindent
Recent CLEO measurments \cite{MA} give the branching ratio 
$BR( {\bar B}^0 \to \pi^+ l {\nu}_l) = 
(1.8\pm0.4\pm0.3\pm0.2)\times10^{-4}$, and $|V_{ub}/V_{cb}| = 0.080\pm0.015$,
assuming $V_{cb}= 0.0381$. %However, $f_+^{B\pi} (q^2)$ 
%is very difficult to measure. 
Using these data we can make predictions 
for  the other decay widths relative to 
$B^0 \to \pi^- l {\nu}_l$. 
In Table II we present the predictions for the ratios 
$\Gamma(B \to P l \nu_l)/ \Gamma(B^0 \to \pi^- l \nu_l)$, 
where B is either a $B^+$ or $B_s^0$ meson  and P is a light 
pseudoscalar meson, for all three values of the 
$\eta - {\eta}^{\prime}$ mixing angle $\theta$. 
Note that we also have assumed 
the vector meson pole dominance as in (\ref{e20}).

We analysed the effects  of $SU(3)$ flavor 
symmetry breaking on the se\-mi\-lep\-to\-nic decays of heavy to 
light pseudoscalar mesons.
The ratios of the matrix elements of local quark  
operators, which transform as triplets, by definition under 
flavor $SU(3)$, 
were estimated using the Wigner-Eckart theorem 
for the {\it state vectors}, but 
including the effects of the substantial mass splitings.   
Using experimental data for $f_+^{DK}(0)$, $|f_+^{D\pi}(0)/f_+^{DK}(0)|$, 
and $f_{D_s}$ we predict $f_D = 209 \pm 39$ MeV, in good agreement with 
the lattice calculations \cite{GM}.

We also determined the ratios  $f_+^{D_s \eta (\eta^{\prime})}(0)/f_+^{DK}(0)$ 
assuming that the form factors $f_+(q^2)$ are pole-dominated. 
We tested our approach using the 
$D_s \to \eta (\eta^{\prime}) l \nu_l$ decay rates and verified its 
consistency with previous analyses.

Within the flavor triplet of $B^+,B^0$, and $B_s^0$ we 
calculated the ratios 
$f_{B_s}/f_B = 1.04$, and $f_+^{B\pi}(0)/f_+^{BK}(0) = 1.08$, 
which agree well with previous estimates. 
Finally, the ratio of decay widths 
$\Gamma (B \to P l \nu_l)/ \Gamma(B^0 \to \pi^- l \nu_l)$ 
were predicted.

We emphasize that assuming only that the meson {\it state vectors} 
are not appreciably affected by flavor $SU(3)$ breaking allows the effects 
of the substantial mass splitings due to flavor $SU(3)$ breaking to be taken 
into account very straightforwardly, 
leading to several  relations among D and B meson 
decay form factors, which are in good agreement with the data. 

\vskip 0.5cm
%{\it Acknowledgement.} 
This work was supported in part by the
Ministry of Science and Technology of the Republic
of Slovenia (B.B. and S.F.), by the British Royal Society (B.B.) 
and by the U.S. Department
of Energy, Division of High Energy Physics,
under grant No. DE-FG02-91-ER4086 (R.J.O.).
S.F. thanks  the Department of Physics and 
Astronomy at Northwestern University, for very kind hospitality.
She also thanks D. Be\' cirevi\' c for many very useful 
discussions.

\newpage
\begin{table}[h]
\begin{center}
\begin{tabular}{|c||c|c|}
\hline
$\theta$ & $D_s \to \eta \mu \nu_{\mu}$ & 
$D_s \to \eta^{\prime} \mu \nu_{\mu}$ \\
\hline
\hline
$-10^{0}$ & $ 2.1 \cdot 10^{-2}$& $ 5.8 \cdot 10^{-3}$ \\
$-20^{0}$ & $ 1.4 \cdot 10^{-2}$& $ 7.7 \cdot 10^{-3}$ \\
$-23^{0}$ & $ 1.1 \cdot 10^{-2}$& $ 8.2 \cdot 10^{-3}$ \\
\hline\hline
$exp.$ & $ (3.3 \pm1.0) \%$& $ (8.7\pm3.4) \cdot 10^{-3}$ \\
\hline
\end{tabular}
\end{center}
\caption{The branching ratios for $D_s \to \eta \mu \nu_{\mu}$ 
and $D_s \to \eta^{\prime} \mu \nu_{\mu}$ decays calculated for 
typical values of the $\eta - \eta^{\prime}$ 
mixing angle $\theta$.}
\end{table}

\begin{table}[h]
\begin{center}
\begin{tabular}{|c||c|}
\hline
$ B\to P$ & $\Gamma (B \to P l \nu_l)/ \Gamma(B^0 \to \pi^- l \nu_l)$\\
\hline
\hline
$B^+ \to \pi^0$ & $ 0.5$ \\
\hline
$B^+ \to \eta (\theta = -10^0)$ & $ 0.21$ \\
$B^+ \to \eta (\theta = -20^0)$ & $ 0.29$ \\
$B^+ \to \eta (\theta = -23^0)$ & $ 0.31$ \\
\hline
$B^+ \to \eta^{\prime} (\theta = -10^0)$ & $ 0.16$ \\
$B^+ \to \eta^{\prime} (\theta = -20^0)$ & $ 0.11$ \\
$B^+ \to \eta^{\prime} (\theta = -23^0)$ & $ 0.09$ \\
\hline
$B_s^0 \to K^-$ & $ 0.92$\\
\hline
\end{tabular}
\end{center}
\caption{The ratios  $\Gamma (B \to P l \nu_l)/ 
\Gamma(B^0 \to \pi^- l \nu_l)$. }
\end{table}

\end{document}